\documentclass[11pt]{article}
\usepackage[top=1in,left=1in,right=1in,bottom=1in]{geometry}

\usepackage{multirow,times}

\usepackage{amsmath,amsthm,amssymb,graphicx}

\usepackage{apalike}
\usepackage{breakcites}

\usepackage[colorlinks=true,urlcolor=red, 
linkcolor = blue,citecolor=blue]{hyperref}
\usepackage[format=plain,labelfont=it,
    textfont=it]{caption} 

\theoremstyle{remark}
\newtheorem*{Remark}{Remark}

\begin{document}

\title{\bf Quantifying the common genetic variability of bacterial traits}

\author{T. Tien Mai$^{(1)}$ , Gerry Tonkin-Hill$^{(2),(3)}$  , John A Lees$^{(4),(6)}$
\\
and Jukka Corander$^{(2),(3),(5)}$ }

\date{
\footnotesize
$^{(1)}$ Department of Mathematical Sciences, Norwegian University of Science and Technology, Trondheim, Norway.
\\
$^{(2)}$ Parasites and Microbes, Wellcome Sanger Institute, Cambridgeshire, UK.
\\
$^{(3)}$ Oslo Centre for Biostatistics and Epidemiology, 
Department of Biostatistics, University of Oslo, Norway.
\\
$^{(4)}$ MRC Centre for Global Infectious Disease Analysis, Department of Infectious Disease Epidemiology, Imperial College, UK.
\\
$^{(5)}$ Department of Mathematics and Statistics,
 University of Helsinki, Finland. 
 \\
 $^{(6)}$European Molecular Biology Laboratory, 
 European Bioinformatics Institute EMBL-EBI, 
 Hinxton, UK.
\vspace*{.1cm}
\\
email: the.t.mai@ntnu.no
}

\maketitle

\begin{abstract}
The study of common heritability, or co-heritability, among multiple traits has been widely established in quantitative and molecular genetics. However, in bacteria, genome-based estimation of heritability has only been considered very recently and no methods are currently available for considering co-heritability. Here we introduce such a method and demonstrate its usefulness by multi-trait analyses of the three major human pathogens \textit{Escherichia coli}, \textit{Neisseria gonorrhoeae} and \textit{Streprococcus pneumoniae}. We anticipate that the increased availability of high-throughput genomic and phenotypic screens of bacterial populations will spawn ample future opportunities to understand the common molecular basis of different traits in bacteria.    
\end{abstract}

\section{Introduction}
Understanding heritability of traits is one of the cornerstones of both quantitative and molecular genetics, representing a rich history that spans more than a century of research \cite{LynchWalshbook}, \cite{BurgerBook}. More recently, the genetic relationships between different traits have received substantial attention in human molecular genetics, for example leading to the identification of important relationships such as the one between height and coronary artery disease, as well as between ulcerative colitis and childhood obesity \cite{bulik2015atlas}. While methods for genome-wide association studies (GWAS) originally developed in human genetics have been successfully adapted to the field of bacterial genomics  \cite{Lees2016-gv,Collins2018-jj,Earle2016-ro}, genome-wide estimation of heritability in bacteria has only more recently received  attention \cite{Lees2017-js,Mai2021-ue}. Thus far, identification of genetic correlations between traits in bacteria has received little attention. Hence, there is an opportunity to broaden our understanding of the genetic correlations of important bacterial traits, including antibiotic resistance, transmission rate, carriage duration, virulence and gene expression, to name a few. 

Compared to studies in human genetics, genetic association studies in bacteria present unique methodological challenges due in part to the extensive variation in gene content and sequence diversity even within members of the same species \cite{Tettelin2005-zx}. The clonal reproduction mode of bacteria further implies that linkage disequilibrium (LD) typically stretches much further than in sexually reproducing species and consequently many genomic regions may be associated with any given trait \cite{Chen2015-tz}. These issues have been addressed through the use of short reference independent sequence features as a replacement for the SNP-based analyses common in human genetics \cite{Lees2016-gv,Jaillard2018-ge}. Simulations, linear mixed models (LMMs) and penalised multiple regression have also been used to account for the highly clonal nature of bacterial genome datasets \cite{Lees2016-gv,Collins2018-jj,Lees2020-zt}. Lineage level effects are often reported when very high LD obscures locus-specific associations \cite{Earle2016-ro}.

These advances have led to significant discoveries between genetic variation in bacteria and important traits such as antibiotic resistance \cite{Earle2016-ro}, carriage duration \cite{Lees2017-js} and virulence \cite{Chaguza2020-hr,Weinert2015-vu}. However, investigations of genetic correlations between such traits have so far mostly been restricted to broader scale comparisons of lineages, as exemplified by the identification of an association between antimicrobial resistance and carriage duration in \textit{Streptococcus pneumoniae} \cite{Lehtinen2017-em}.

There exists a large body of literature for investigating genetic correlations between traits in humans and in animals in general. These include those based on Mendelian randomisation, which focuses on investigating previous variants found to be significantly associated with a phenotype, to investigate causal relationships between risk factors and disease. Other approaches include restricted maximum likelihood (REML) and polygenic scores which typically make use of linear mixed models (LMMs) to account for all observed SNPs. A comprehensive discussion of these approaches can be found in Van Rheenen et al., \cite{Van_Rheenen2019-bp}. For many complex human traits such as height, heritability is spread over many sites with each contributing a small effect \cite{Yang2010-qh,speed2017reevaluation,speed2019sumher}. Conversely, in bacteria, many phenotypes of interest are the result of a small number of variants with larger effect sizes, with a notable example being the resistance conferring mutations in the penicillin binding proteins of \textit{S. pneumoniae} \cite{Chewapreecha2014-mj}. For this reason, in bacterial genomics, penalised regression techniques present an attractive alternative to the more common LMM approach as it has been shown to better account for the clonal population structure of bacteria \cite{Saber2020-ge,Lees2020-zt}.

Here, we develop a high-dimensional penalised linear regression model to investigate genetic correlations between bacterial phenotypes. We demonstrate the accuracy and advantages of the approach by investigating the association between resistance to different antibiotic classes, carriage duration and virulence in a selection of \textit{Streptococcus pneumoniae}, \textit{Escherichia coli} and \textit{Neisseria gonorrhoeae} data sets.

\section{Results}
\label{sc_result}

\subsection{Overview}

To investigate the genetic covariance and coheritability between two phenotypic traits of interest, we built off of previous work which used elastic net regression to generate accurate phenotypic prediction models and more accurately estimated heritability in related populations \cite{Lees2020-zt}. Here, we model each phenotype as a linear combination of genetic variants with independent error terms to account for environmental and unmeasured genetic effects. The genetic covariance between the two traits is then obtained by subtracting the covariance of the inferred error terms from the covariance between the two traits of interest (Methods). This can then be standardised to obtain the genetic correlation with the resulting scale being between -1 indicating perfect negative correlation and 1 indicating perfect positive correlation between the estimated genetic effects. Similar to previous studies in human genetics, in addition to genetic correlation, we also consider the covariance of the causal effects \cite{Shi2017-jv,bulik2015ld}. This allows for the separation of causal correlations from those that may be driven by linkage disequilibrium (LD) and is particularly important in analysing highly clonal datasets such as those commonly found in bacterial genomics. Using this approach, we examined the genetic correlations between resistance to different antibiotics across a number of bacterial species as well as virulence and carriage duration in \textit{Streptococcus pneumoniae}.

\subsection{Antibiotic resistance, virulence and carriage duration are genetically correlated in \textit{Streptococcus pneumoniae}}

\begin{figure}
	\centering
	\includegraphics[scale=.8]{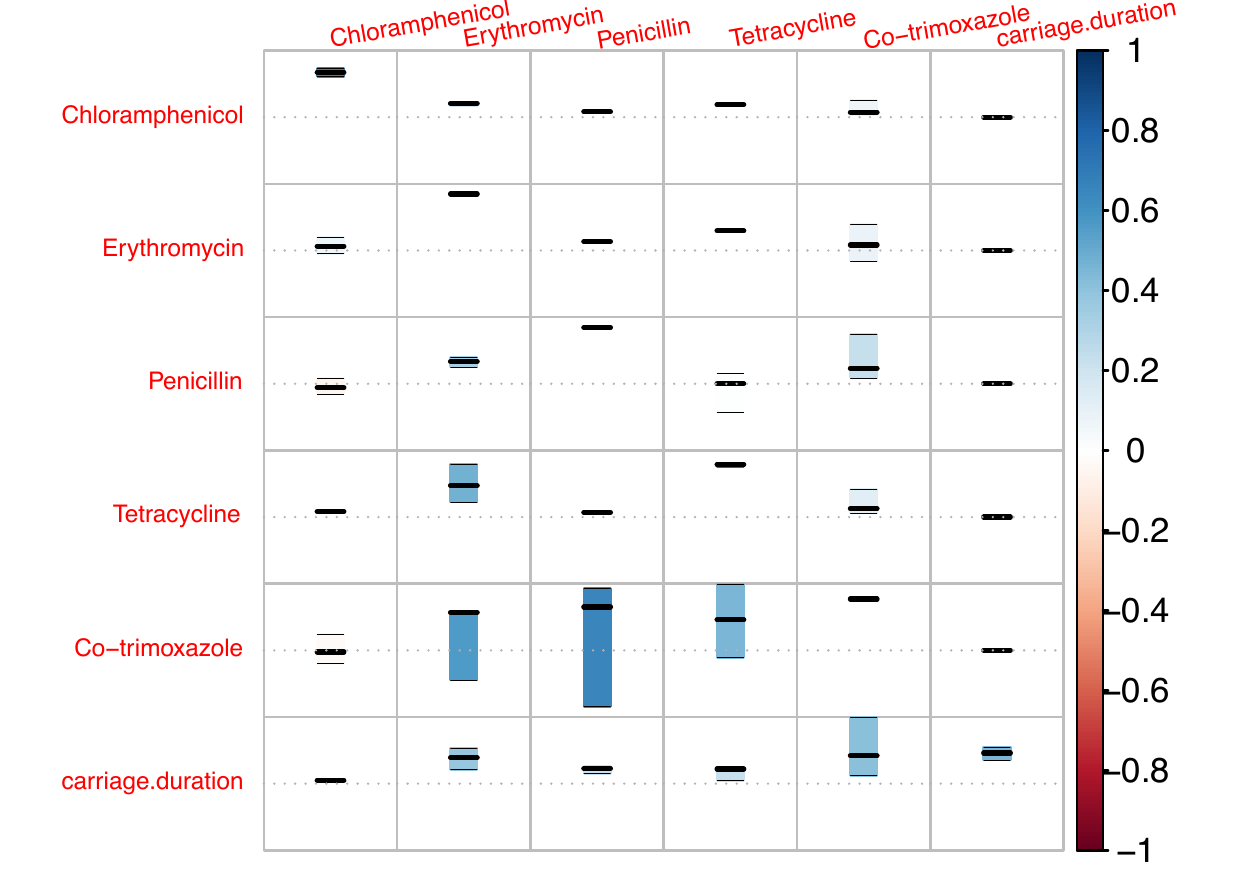}
	\caption{\textbf{Maela} data. Heritability in diagonals, genetic relatedness in the upper diagonals, genetic correlation in lower diagonals.}
	\label{fg_maela}
\end{figure}

To examine the ability of our approach to identify meaningful genetic correlations between bacterial phenotypes, we initially considered two large \textit{S. pneumoniae} datasets. The Maela dataset consists of 3069 whole genomes taken from an infant cohort study in a refugee camp on the Thailand-Myanmar border \cite{Chewapreecha2014-kn}. This dataset has been used in genome wide association studies to identify genetic loci associated with antibiotic resistance pneumococcal carriage duration \cite{Lees2017-js,Chewapreecha2014-mj}. Using our approach, we investigated the genetic correlation between the resistance to five antibiotics and carriage duration. Figure \ref{fg_maela} indicates that pneumococcal carriage duration is positively correlated with resistance particularly in the case of Erythromycin. We found a similar correlation in the cases of Penicillin, Tetracycline and co-trimoxazole although there was a higher degree of uncertainty in these estimates. This is consistent with a recent study that investigated the association between resistance and carriage duration in the Maela dataset \cite{Lehtinen2017-em}. Interestingly, we found that there appears to be no covariance between the causal effects of carriage duration and resistance to any of the antibiotics considered. While laboratory studies have shown that resistance to antibiotics leads to reduced growth rates in \textit{S. pneumoniae} \cite{Rozen2007-qc,Trzcinski2006-ir} these results suggest that it is not the fitness costs of resistance that drive the association with carriage duration. Rather the fitness benefits of resistance are likely to be greater for those lineages with a longer duration of carriage due to the increased probability that they encounter antimicrobial treatment.

\begin{figure}[!t]
	\centering
	\includegraphics[scale=.7]{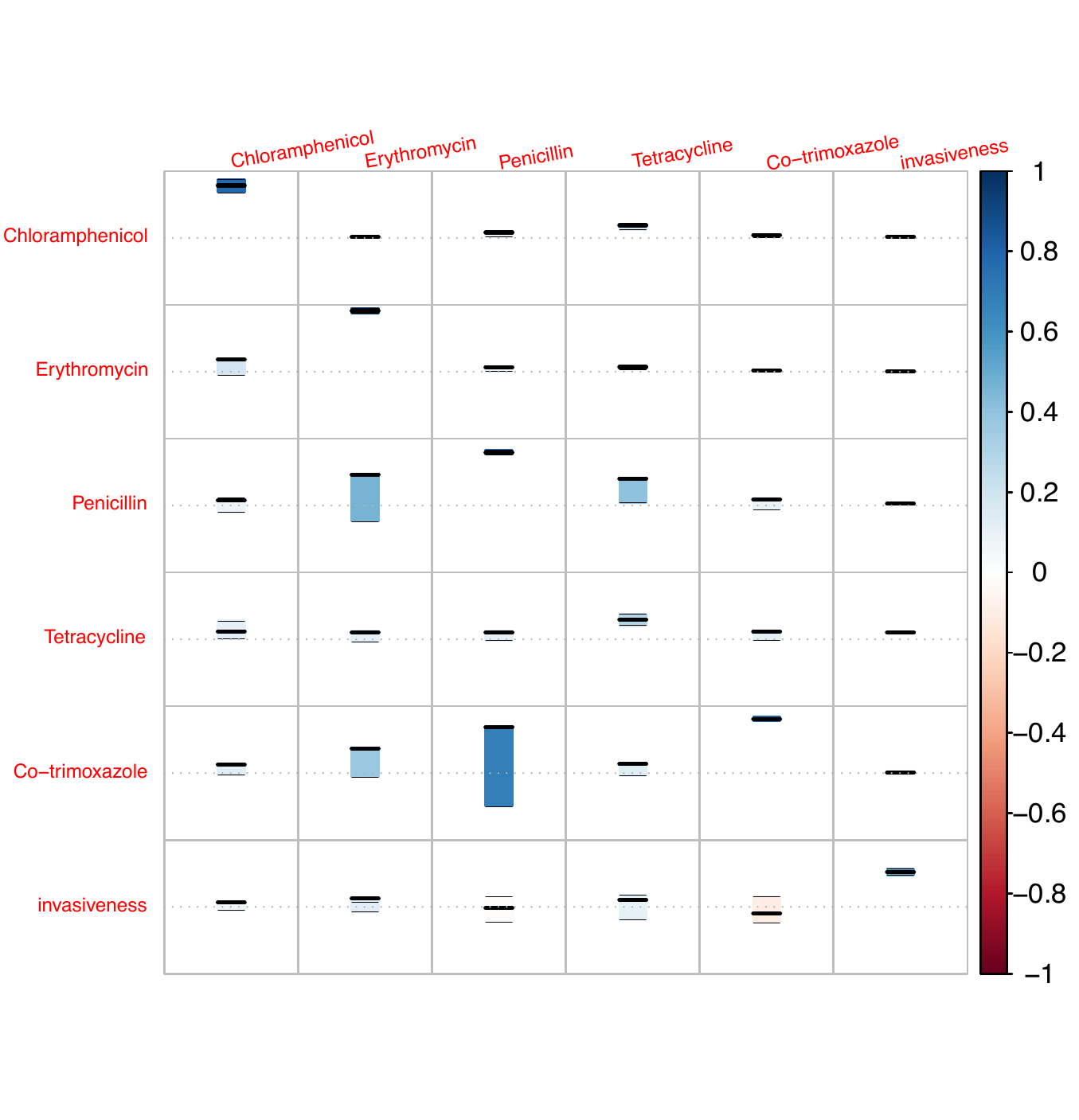}
	\caption{\textbf{Streptococcus pneumoniae data}. Heritability in diagonals, genetic relatedness in the upper diagonals, genetic correlation in lower diagonals.}
	\label{fg_SP}
\end{figure}

In addition to the correlation with carriage duration, we found that there were strong genetic correlations between resistance to each of the antibiotics considered (Figure \ref{fg_maela}). In the cases of tetracycline resistance, erythromycin resistance, and chloramphenicol resistance, we also observed correlations between the causal effects. To investigate whether these correlations were unique to this dataset, we ran the same analysis on a large subset of the Global Pneumococcal Sequencing project consisting of genomes sampled in South Africa for which accurate resistance and virulence phenotypic information was known \cite{Lo2019-gh,Gladstone2019-rz}. Figure \ref{fg_SP} shows that a very similar profile of genetic correlations between resistance to the same antibiotics was present in this dataset, which indicates that our results are consistent across multiple distinct locations. Although resistance to these antibiotics is thought to be mediated by different mechanisms, the corresponding resistance genes can be found in a single cassette ICESp23FST81. This cassette can be inherited both vertically and horizontally and varies in which genes it harbours: \textit{tetM} for tetracycline; \textit{ermB} or \textit{mel} and \textit{mef} for erythromycin; \textit{cat} for chloramphenicol \cite{Croucher2011-fy}. Thus, the causal correlation identified is likely to be driven by the presence and absence of this cassette which is not in strong LD with the rest of the genome.

The South African dataset also included accurate information on which genomes were isolated from invasive disease cases allowing us to consider the genetic correlation between resistance and virulence. We found the virulence was negatively genetically correlated with resistance although there was considerable uncertainty in these estimates. Similar to resistance and carriage duration, we did not observe a strong covariance between the causal effects of virulence and resistance. The anti-correlation between virulence and resistance has been observed previously in isolates of \textit{S. pneumoniae} from China where resistance was observed at much lower frequencies in isolates taken from children with meningitis \cite{Wang2019-sj}. A similar finding has also been made in \textit{Klebsiella pneumoniae} where common virulence genes are rarely found together with antimicrobial resistance genes \cite{Holt2015-ql}. A potential explanation for this is that, in contrast to carriage duration, resistance is less beneficial to those pneumococcal lineages that commonly cause invasive disease. In \textit{S. pneumoniae}, these lineages are often rarely found in carriage studies indicating they may encounter antimicrobial treatment less frequently than the common multidrug resistant lineages \cite{Chaguza2016-la,Gladstone2019-rz}.

\subsection{Genetic correlation between resistance to different antibiotics is observed in multiple bacterial species}

\begin{figure}
	\centering
	\includegraphics[scale=.7]{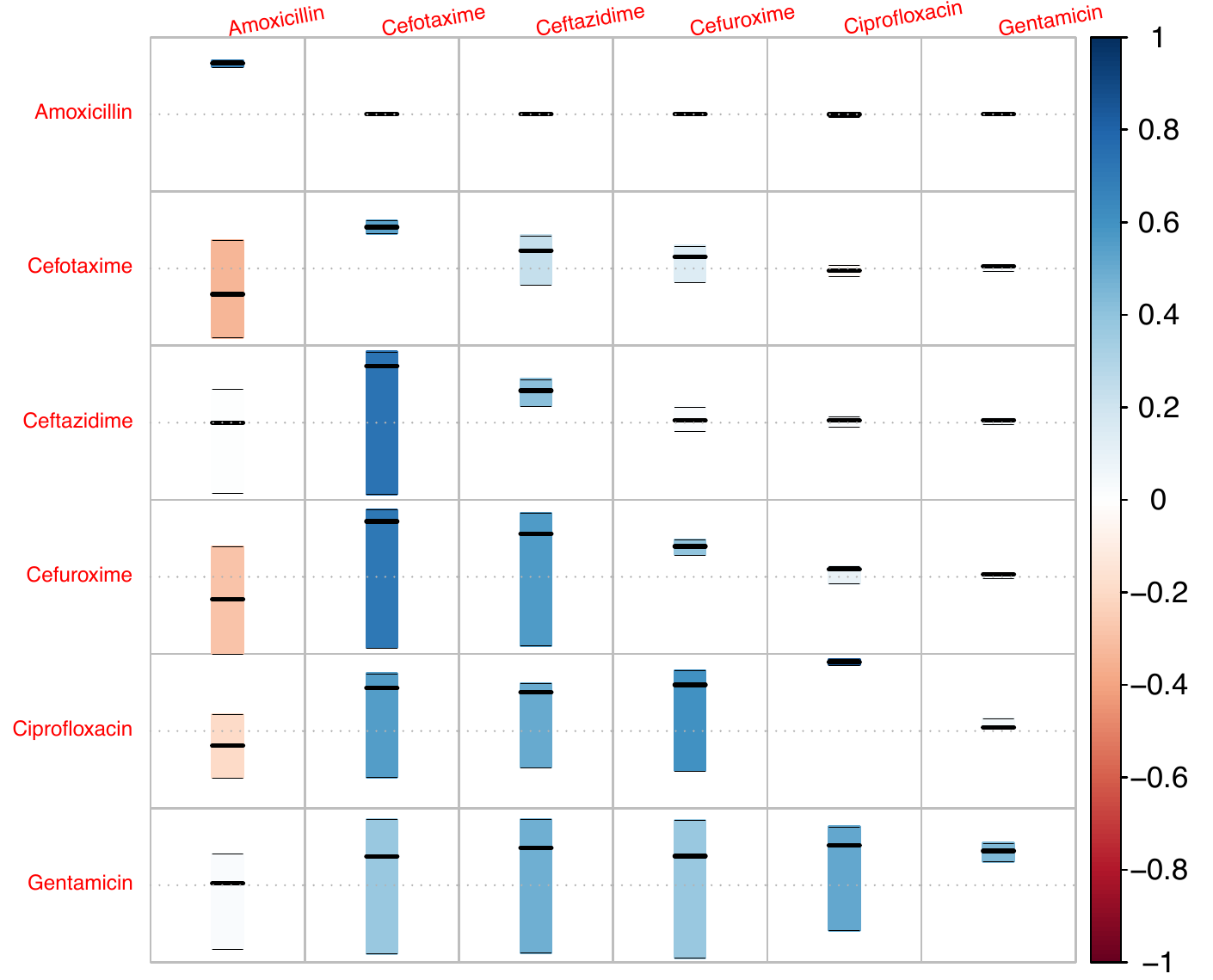}
	\caption{\textbf{Ecoli data}. Heritability in diagonals, genetic covariance in the upper diagonals, genetic correlation in lower diagonals.}
	\label{fg_ecoli}
\end{figure}

We also observed genetic correlations between resistance phenotypes in both \textit{Escherichia coli} and \textit{Neisseria gonorrhoeae}. Using a dataset of 1509 \textit{Escherichia coli} isolates taken from an 11 year systematic hospital survey of bacteremia associated isolates in England, we explored whether there were identifiable correlations between the resistance profiles of each isolate \cite{Kallonen2017-ar}. Figure \ref{fg_ecoli} demonstrates that with the exception of Amoxicillin there was a positive genetic correlation between the resistance phenotypes. However, we observed considerable uncertainty in these estimates and thus can not rule out correlations between the other resistance profiles. 

\begin{figure}
	\centering
	\includegraphics[scale=.7]{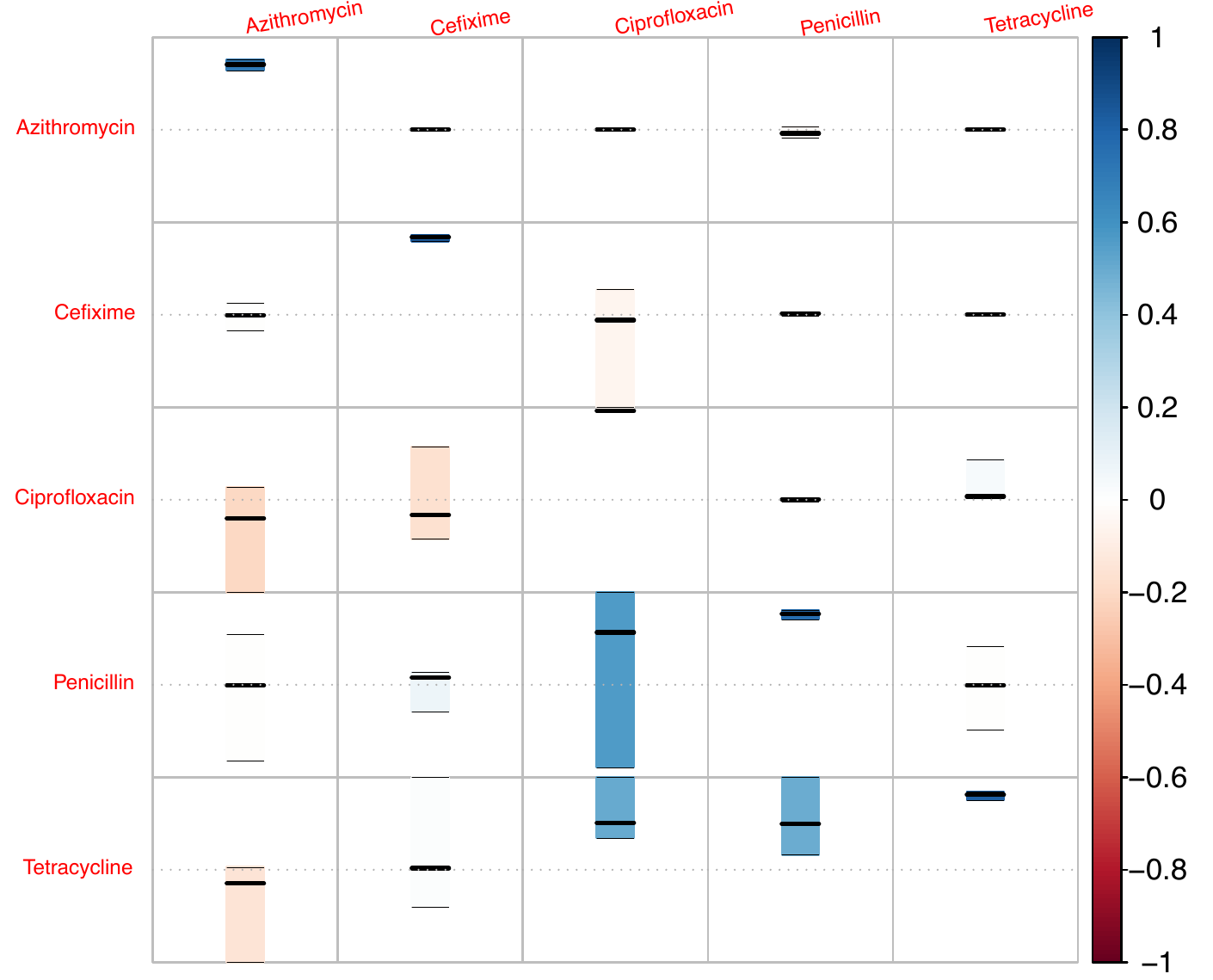}
	\caption{\textbf{Neisseria gonorrhoeae data}. Heritability in diagonals, genetic covariance in the upper diagonals, genetic correlation in lower diagonals.}
	\label{fg_ng}
\end{figure}

We also investigated the genetic correlations between antibiotic resistance phenotypes in 1595 \textit{N. gonorrhoeae} isolates collected from the USA, Canada and England \cite{Unemo2016-kf,Grad2016-pc,Demczuk2015-px,De_Silva2016-xs,Schubert2018-ux}. This identified strong positive genetic correlations between the resistance phenotypes to Tetracycline, Ciproflaxcin and Penicillin (Figure \ref{fg_ng}). Similar to \textit{S. pneumoniae}, \textit{N. gonorrhoeae} is thought to have different resistance mechanisms to each of these antibiotics. However, unlike in \textit{S. pneumoniae} we did not find a causal association between the effects of each antibiotic. Rather it is likely that the genetic correlation is driven by the two major modern gonococcal lineages with a multidrug resistance lineage being common in high-risk sexual networks and a multisusceptible lineage often associated with lower risk heterosexual networks \cite{Sanchez-Buso2019-kv}. Strikingly, we also observed a negative genetic correlation between azithromycin and tetracycline or ciprofloxacin albeit with a greater level of uncertainty. This is consistent with the very rare instances of resistance to the common dual therapy (injectable ceftriaxone plus oral azithromycin) suggesting that the fitness costs of gaining resistance to azithromycin in addition to certain other antibiotics is sufficiently high to prevent such isolates from proliferating \cite{Sanchez-Buso2019-kv,Grad2016-pc}. 

\section{Discussion}

\label{sc_discussion}

The increasingly common use of genome wide association studies in bacterial genomics has led to vast improvements in our understanding of the links between bacterial genomes and important phenotypes such as antibiotic resistance and virulence. Here, we have developed a method to estimate the coheritability between traits in bacterial populations. Building on our earlier work using elastic net regression to fit whole genome models to estimate regression slopes \cite{Lees2020-zt}, and our approaches to more accurately estimate heritability in related populations \cite{Mai2021-ue}, we are able to use the covariance between fitted predictor values to estimate shared genetic contributions between pairs of traits. By looking at causal correlations, we have demonstrated that it is possible to distinguish when two phenotypes are likely to be the result of similar genetic mutations or when population structure and epidemiology could be driving the genetic correlation. This is highlighted by the observed associations between resistance, virulence and carriage duration in \textit{S. pneumoniae}. Here, statistical genetics approaches are particularly powerful as these phenotypes can be difficult to measure in lab models, whereas they can be more readily quantified in the natural population. We show that consistent with their known causal genetic mechanisms, typically point mutations or genes which are not co-localised, these phenotypes are unlikely to be driven by the same genetic features.

In \textit{S. pneumoniae} we found evidence of co-heritability between tetracycline resistance, erythromycin resistance, and chloramphenicol resistance -- though the strength and significance of this finding varied between populations. Although the causal elements themselves are not directly coheritable, the molecule they can be found on is inherited both vertically and horizontally. From a more methodological perspective, the SNPs on these elements would typically all be associated with each phenotype, and shared when multiple resistances are present \cite{Lees2016-gv}. The replication of results across distinct pneumococcal datasets in Thailand and South Africa suggests that the method is robust to the impacts of population structure and sampling biases. We also demonstrated weaker genetic correlations between resistance to antibiotics in \textit{N. gonorrhoeae} and \textit{E. coli}. Unlike in \textit{S. pneumoniae}, these were found to not have causal correlations and are likely to be the result of the significant resistance associated with population structure observed in the species.

In summary, our method is capable of quantifying the shared genetic variability related to variance in pairs of phenotypes, in closely related bacterial populations. We did not explicitly model LD or relatedness, nor look at accessory genome variation or more complex genetic variation. Additionally, our findings at this point are largely observational. Nevertheless, we believe this will be a useful first step into further research on shared genomic bases between traits in pathogenic bacteria. To enable other researchers to easily conduct similar analyses, we have also provided an accompanying R package on GitHub (\url{https://github.com/tienmt/coher}).

\section*{Acknowledgments}
TTM is supported by the Norwegian Research Council grant number 309960 through the Centre for Geophysical Forecasting at NTNU. JAL acknowledges funding from the MRC Centre for Global Infectious Disease Analysis (reference MR/R015600/1), jointly funded by the UK Medical Research Council (MRC) and the UK Foreign, Commonwealth \& Development Office (FCDO), under the MRC/FCDO Concordat agreement and is also part of the EDCTP2 programme supported by the European Union.


\appendix

\section{Methods}
\label{sc_methods}
\subsection{Model and definitions}
Given two traits $ y $ and $ z $, each is modelled as a linear combination of $ p $ genetic variants $ X_{.j} $ and an error term (environmental and unmeasured genetic effects)
\begin{equation}
\begin{aligned}
\label{linear.model}
y_{n} = X_{n\times p} \beta_{p} + \varepsilon_{n},
\\
z_{n} = X_{n\times p} \alpha_{p} + \gamma_{n}.
\end{aligned}
\end{equation}
We assume that $ X_{i.} $ are i.i.d random vector with 0-mean and covariance matrix $\Sigma$ and that the random noises  $ \varepsilon_{i}, \gamma_{i} $ are independent of $ X $ with 0-mean and the variances $ \sigma_{\varepsilon}^2, \sigma_{\gamma}^2 $. We assume that $ y $ and $ z $ have 0-mean.

\begin{Remark}
	In this work, we do not assume the effects $\beta, \alpha $ follow any distribution. This is different with most works in the literature that conducted in linear mixed model. In linear mixed model, the elements of $\beta$ (and $\alpha $) are considered as i.i.d random variables following a Gaussian distribution i.e $\beta_j \overset{i.i.d}{\sim} \mathcal{N} (0, \sigma^2_{\beta}) $, while the genetic covariates $X$ are assumed fixed.  
\end{Remark}

\subsubsection*{Genetic covariance and correlation}
From the problem formulation \eqref{linear.model}, the covariance between the phenotypes $y $ and $z$ can be explained as a summation of the genetic covariance and environmental covariance as follow
\begin{align*}
{\rm Cov}(y,z) 
& =
\mathbb{E} (yz) - \mathbb{E} (y)\mathbb{E} (z)
\\
& =
\mathbb{E} \left\langle  X\beta + \varepsilon , X\alpha + \gamma \right\rangle
\\
& =
\alpha^\top \mathbb{E} \left( X^\top X \right)\beta  + \mathbb{E}(\varepsilon \gamma)
\\
& =
\alpha^\top \Sigma \beta + {\rm Cov}(\varepsilon, \gamma).
\end{align*}
As a consequence, the \textit{genetic covariance} between two traits $y$ and $z$ is defined by
\begin{align}
\boxed{ {\rm g.Cov}(y,z) = \alpha^\top \Sigma \beta },
\end{align}
and the \textit{genetic correlation} of two traits $y$ and $z$ is defined by 
\begin{align}
\label{genetic.correlation}
\boxed{ 
	{\rm g.Cor}(y,z) =  \frac{ \alpha^\top \Sigma \beta }{ \sqrt{\alpha^\top \Sigma \alpha }
		\sqrt{ \beta^\top \Sigma \beta }  } 
},
\end{align}
which is a standardization of the genetic covariance.

The quantities $\beta^\top \Sigma \beta $ and $\alpha^\top \Sigma \alpha $ are the genetic variance of each trait and they are used to define the heritability of the trait.

\begin{Remark}
	Our formula in the definition of the genetic correlation \eqref{genetic.correlation} is different to the one defined in \cite{Shi2017-jv} where they define the denominator as the co-heritability
	. Our formula guarantees that the genetic correlation varies between -1 and 1 whereas \cite{Shi2017-jv} have to threshold it.
\end{Remark}

\subsubsection*{Heritability}
Heritability is an important quantitative concept that measures the variation observed in a trait contributed by the variation of genotype. From \eqref{linear.model}, we have
\begin{align*}
{\rm Var}(y_i) 
=  \beta^\top \Sigma \beta  + \sigma_{\varepsilon}^2; \quad
{\rm Var}(z_i) 
=  \alpha^\top \Sigma \alpha + \sigma_{\gamma}^2.
\end{align*}
The (narrow-sense) heritabilities of $y$ and $z$ are defined as
\begin{align}
\label{herit.fomula}
\boxed{
	h^2(y)  = 
	\frac{  \beta^\top \Sigma \beta  }{ \beta^\top \Sigma \beta  + \sigma_{\varepsilon}^2 } 
	; \quad
	h^2(z)  = 
	\frac{ \alpha^\top \Sigma \alpha }{ \alpha^\top \Sigma \alpha + \sigma_{\gamma}^2 }.
}
\end{align}

\subsubsection*{Genetic relatedness}
The genetic relatedness as in~\cite{bulik2015atlas}, or the covariance of the causal effect as in \cite{Shi2017-jv}, between $ y $ and $ z $ is defined as the inner product of the regression coefficients
\begin{align}
\label{gene.related}
\boxed{ {\rm g.R}(y,z) = \left\langle \beta, \alpha  \right\rangle }.
\end{align}
We further consider the normalization of the genetic relatedness as it helps to compare across the phenotypes as well as across the region, it is defined as
\begin{equation}
\boxed{ {\rm g.R}(y,z) = \frac{ \left\langle \beta, \alpha  \right\rangle }{\|\alpha\| \|\beta\|} }.
\end{equation}

\begin{Remark}
	Note that in the special case where there is no LD (i.e. the covariance matrix $\Sigma$ is the identity matrix), the genetic relatedness is exactly the genetic covariance: $ {\rm g.Cov}(y,z) =  \beta^\top \Sigma \alpha = \beta^\top \mathbf{I} \alpha =\left\langle \beta, \alpha  \right\rangle = {\rm g.R}(y,z)  $. To make it more clear, consider a toy example with only 2 genetic variants, let $\beta = (1,0)^\top $ and $\alpha = (0,1)^\top $ be the causal effects of the 2 phenotypes $y$ and $z$ and 
	$$
	\Sigma = \begin{bmatrix}
	1       & 0.8  
	\\
	0.8       & 1  
	\end{bmatrix}.
	$$
	We obtain ${\rm g.R}(y,z) = \left\langle \beta, \alpha  \right\rangle  = 0$ while ${\rm g.Cov}(y,z) =  \beta^\top \Sigma \alpha = 0.8 $. Therefore, when the causal variants are completely distinct for 2 traits, the genetic relatedness is always zero, while the genetic covariance can be different from zero due to the LD. This has also been discussed in \cite{Shi2017-jv}.
\end{Remark}

\subsection{Estimation methods}

A naive approach to calculate genetic covariance and correlation is to first estimate the effect sizes $ \beta,\alpha $ and the covariance matrix $\Sigma$ then plug-in these values into their formula. 

\subsubsection*{Estimating the effect sizes}
To estimate the regression coefficients $ \beta$ (and $\alpha $), we use the Elastic net \cite{zou2005regularization}:
\begin{align*}
\hat{\beta}_{enet} := \arg \min_{\beta_0,\beta} \frac{1}{n} \sum_{i=1}^{n}  \ell(y_i,\beta_0+\beta^T x_i) + \lambda\left[0.5(1-\alpha)||\beta||_2^2 + \alpha ||\beta||_1\right],
\end{align*}
where $\lambda > 0$ and $0\leq\alpha\leq 1$ are tuning parameters. Here \(\ell(a ,b)\) is the negative log-likelihood for an observation e.g. for the linear Gaussian case it is \(\frac{1}{2}(a -b)^2\) and for logistic regression it is $ - a \cdot b + \log (1+e^{b})$. 

The elastic net has been shown to be especially useful when the variables are dependent \cite{zou2005regularization} (LD structure), which is the case of genetic data and especially with bacterial genetic data. It is controlled by \(\alpha\) that balances between lasso (\(\alpha=1\)) and ridge (\(\alpha=0\)). The tuning parameter \(\lambda\) controls the overall strength of the penalty and we use 10-fold cross-validation to choose suitable value for $\lambda$. Elastic net  is implemented in the software 'pyseer' \cite{lees2018pyseer} focusing on bacterial GWAS data and has been shown to outperform some state-of-the-art methods in bacterial GWAS \cite{Saber2020-ge,Lees2020-zt}.

To obtain a 95\% confidence intervals, we propose to use a post-selection strategy by fitting the OLS estimator on the selected covariates and obtain the 95\% confidence intervals for the estimated effect sizes. This is followed from \cite{belloni2013least}.

\subsubsection*{Heritability estimation}
Heritability estimates are performed using the Boosting Heritability approach in \cite{Mai2021-ue} where heritability estimation of commonly used methods is improved by using a sample splitting strategy.

\section{Supplementary additional results}
\label{appendix}

\begin{figure}[!ht]
\caption{ Sample \textbf{Correlation matrix} of phenotypes in Maela data using binary (left) and zone size (right) data}
\label{fg_maela_sample_phenotype}
\centering
\includegraphics[scale=0.4]{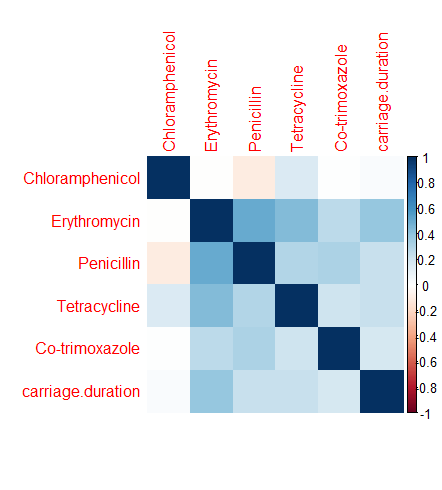}
\includegraphics[scale=0.3]{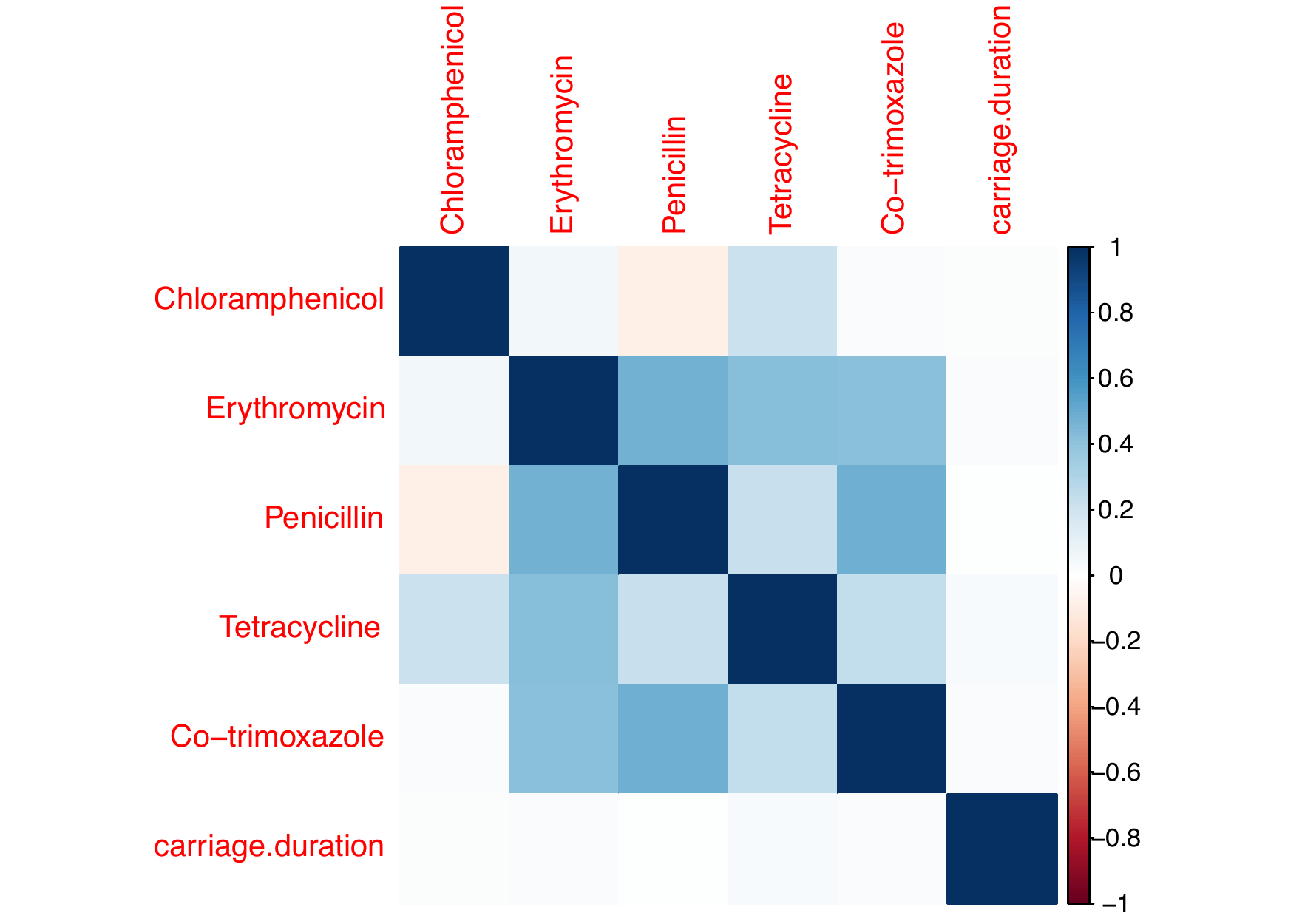}
\end{figure}

\begin{figure}[!ht]
\caption{Histogram of zone size phenotypes in Maela data}
\includegraphics[scale=.4]{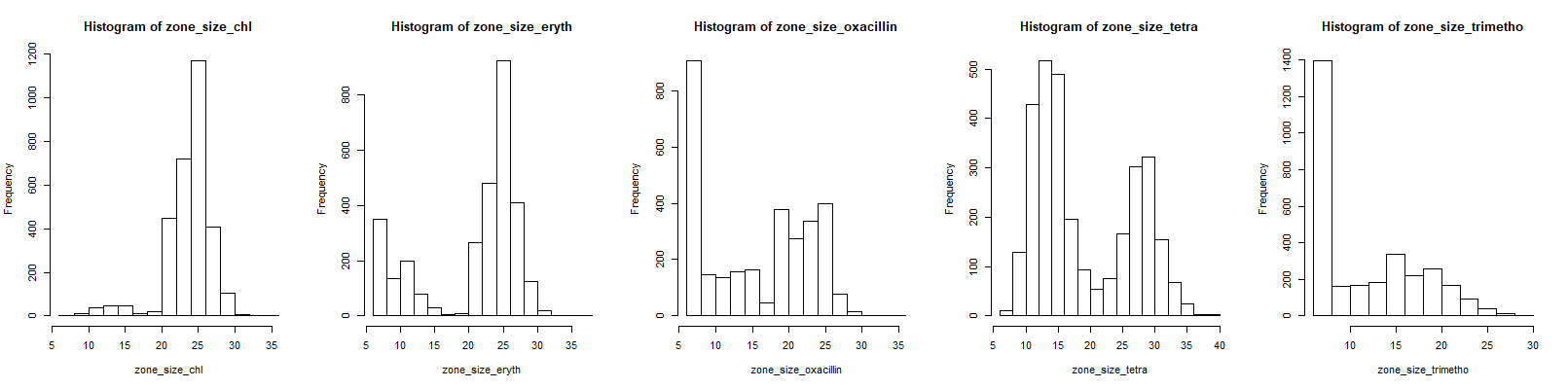}
\end{figure}

\begin{table}
\footnotesize
\centering
\caption{Binary \textbf{Maela} data. Heritability in the diagonal, genetic corelation in lower diagonals, genetic related in upper diagonals}
\makebox[\textwidth]{
\begin{tabular}{| l | c | c| c| c| c| c |}
\hline\hline
	& Chloramphenicol
	& Erythromycin
	& Penicillin
	& Tetracycline
	& Co-trimoxazole
	& carriage.duration 
\\ \hline
\multirow{2}{*}{Chloramphenicol} & \textbf{0.763} & 0 & 0 & 0 & 0 & -0.009
	\\
	& (0.680, 0.838) & (0, 0)  & (0, 0) & (0, 0) & (0, 0) & (-0.091, 0.011) 
		\\ \hline
\multirow{2}{*}{Erythromycin} & -0.017 & \textbf{0.915} & 0.004 & 0 & 0 & 0 
	\\
		& (-0.185, 0.707) & (0.889, 0.941) & (-0.002, 0.009) & (0, 0)& (-0.009, 0.001)& (-0.009, 0.001)
		\\ \hline
\multirow{2}{*}{Penicillin} & -0.108 & 0.586 & \textbf{0.829} & 0 & -0.671 & 0
	\\
		& (-0.326, 0.458) & (-0.999, 0.545)& (0.800, 0.851) & (0, 0) & (-0.999, 0.210) & (0, 0)
		\\ \hline
\multirow{2}{*}{Tetracycline} & 0.025 & 0.441 & 0.261 & \textbf{0.891} & 0.007 & 0
	\\
		& (-0.156, 0.191) & (-0.999, 0.543)& (-0.383, 0.371)& (0.873, 0.908) & (0.001, 0.031) & (0, 0)
		\\ \hline
\multirow{2}{*}{Co-trimoxazole} & -0.127 & 0.448 & 0.571 & 0.284 & \textbf{0.638}& 0
	\\
		& (-0.343, 0.370) & (-0.999, 0.510)& (-0.999, 0.934)& (-0.430, 0.371) & (0.565, 0.709) & (0, 0)
\\ \hline
\multirow{2}{*}{carriage.duration} & 0.029 & 0.436 & 0.241 & 0.253 & 0.299 & \textbf{0.459}	
\\
		& (-0.131, 0.145) & (-0.002, 0.882) & (-0.091, 0.783)& (-0.079, 0.257) & (-0.080, 0.733) & (0.356, 0.547)
\\ 
\hline\hline
\end{tabular}
}
\end{table}

\begin{table}[!ht]
\footnotesize
\centering
\caption{Binary \textbf{Maela} data.  genetic covariance in the lower diagonals, co-variance of causal effects in upper diagonals.}
\makebox[\textwidth]{
\begin{tabular}{| l | c | c| c| c| c| c |}
\hline\hline
	& Chloramphenicol
	& Erythromycin
	& Penicillin
	& Tetracycline
	& Co-trimoxazole
	& carriage.duration 
\\ \hline
\multirow{2}{*}{Chloramphenicol} &  & 0 & 0 & 0 & 0 & -0.080
	\\
	&	& (0, 0) & (0, 0) & (0, 0) & (0, 0) & (-0.541, 0.264)
		\\ \hline
\multirow{2}{*}{Erythromycin} & -0.001 &  & 0.003 & 0 & -0.001 & 0 
	\\
		& (-0.126, 0.115) & () & (-0.001, 0.015) & (0, 0) & (-0.005, 0.001) & (0, 0)
		\\ \hline
\multirow{2}{*}{Penicillin} & -0.010 & 0.095 &   & 0 & -1.501 & 0
	\\
		& (-0.170, 0.144) & (-0.480, 0.831) & () & (0, 0) & (-4.433, 1.021) & (0, 0)
		\\ \hline
\multirow{2}{*}{Tetracycline} & 0.002 & 0.068 & 0.045 &   & 0.015 & 0
	\\
		& (-0.122, 0.124) & (-0.228, 0.477) & (-0.545, 0.923) && (0.005, 0.029) & (0, 0)
		\\ \hline
\multirow{2}{*}{Co-trimoxazole} & -0.009 & 0.055 & 0.078 & 0.037 &  & 0
	\\
		& (-0.816, 0.671) & (-0.686, 0.890) & (-4.227, 5.074) & (-0.782, 1.041) & (0, 0)
\\ \hline
\multirow{2}{*}{carriage.duration} & 0.006 & 0.154 & 0.095 & 0.096 & 0.090 &  	
\\
		& (-0.312, 0.364) & (-0.001, 0.529) & (-0.508, 0.783) & (-0.203, 0.581) & (-0.563,  0.733)
\\
\hline\hline
\end{tabular}
}
\end{table}

\begin{table}
\footnotesize
\centering
\caption{Zone sizes \textbf{Maela} data. (unnormalized) genetic covariance in the lower diagonals,  (unnormalized) covariance of causal effects in upper diagonals. }
\makebox[\textwidth]{
\begin{tabular}{| l | c | c| c| c| c| c |}
\hline\hline
	& Chloramphenicol
	& Erythromycin
	& Penicillin
	& Tetracycline
	& Co-trimoxazole
	& carriage.duration 
		\\ \hline
\multirow{2}{*}{Chloramphenicol} &  & 0.447 & 0.166 & 0.338 & 0.219  & 0
		\\ 
		& & (0.269, 0.657) & (0.099, 0.249) & (0.207, 0.498) & (0.163, 0.282) &(0, 0)
		\\ \hline
\multirow{2}{*}{Erythromycin} &  .026  &  & 0.437 & 0.848 & 0.383 & 0
		\\
		&   (-.0298, .0716)  & & (0.276, 0.634) & (0.589, 1.161) & (-0.228, 1.205) & (0, 0)
		\\ \hline
\multirow{2}{*}{Penicillin} & -0.030  & 0.542  &  &  0.010 & 1.040 & 0
		\\
		&  (-0.109, 0.037) &  (0.225, 0.978) && (-4.418, 4.194) & (0.536, 1.756) & (0, 0)
		\\ \hline
\multirow{2}{*}{Tetracycline} & 0.029  & 0.434  & 0.096  &  & 0.482 & 0
		\\
		&  (0.019, 0.042) & (0.368, 0.507) & (0.056,  0.146) & & (0.324, 0.669) & (0, 0)
		\\ \hline
\multirow{2}{*}{Co-trimoxazole} & -0.008  & 0.728  &  1.056  & 0.501  & 	&  0.031	
\\
&  (-0.378,  0.248) & (-1.375, 3.515) &  (-3.503, 6.090) & (-0.281,  1.499) & 	& (-0.065, 0.142)	
\\ \hline
\multirow{2}{*}{carriage.duration} & 0.017  & 0.391  & 0.290 & 0.187  & 0.423 & 	
\\
&  (-0.050, 0.055) &  ( 0.200, 0.807) & (-1.341, 0.619) &  ( 0.184, 0.651) &  (-2.356, 1.056) & 
\\
\hline\hline
\end{tabular}
}
\label{tb_Maela_appendix}
\end{table}


\begin{figure*}[!ht]
    \centering
    \includegraphics[scale=0.3]{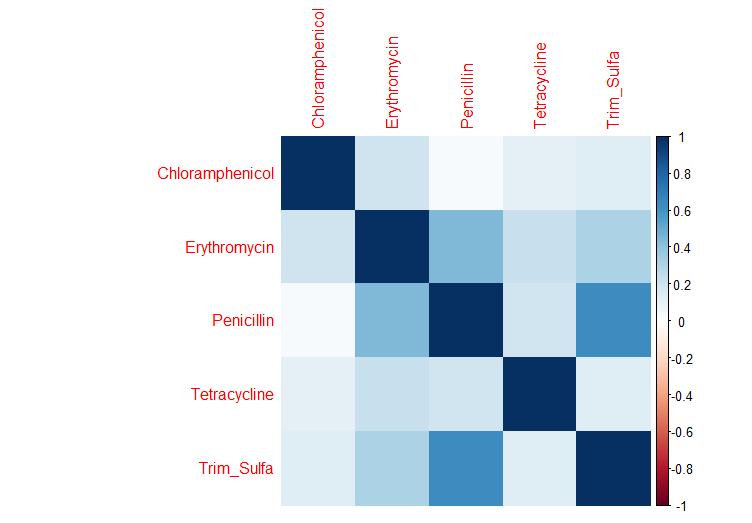}
    \includegraphics[scale=0.3]{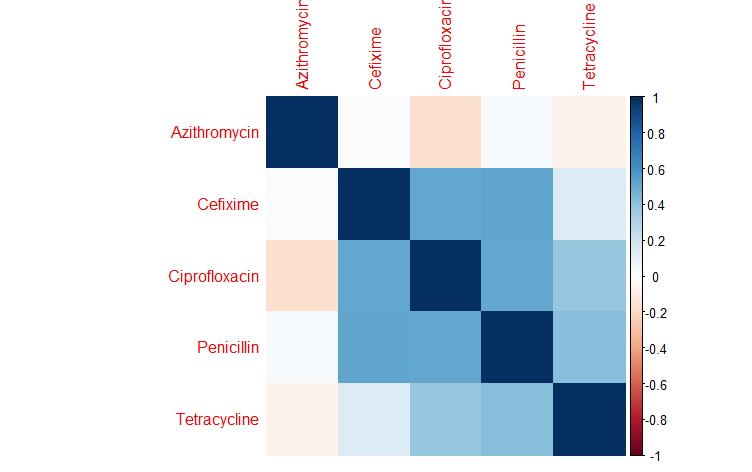}
    \caption{Left: sample correlation matrix of phenotypes in \textbf{Streptococcus pneumoniae} invasiveness data. Right: sample correlation matrix of log(MIC) phenotypes in \textbf{Neisseria gonorrhoeae data} }
    \label{fg_NGdata_sample_phenos}
\end{figure*}

\begin{table}[!ht]
\footnotesize
\centering
\caption{\textbf{Streptococcus pneumoniae data}. genetic covariance in the lower diagonals,  covariance of causal effects in upper diagonals. }
\makebox[\textwidth]{
\begin{tabular}{| l | c | c| c| c| c| c |}
\hline\hline
	& Chloramphenicol
	& Erythromycin
	& Penicillin
	& Tetracycline
	& Co-trimoxazole
	& invasiveness
\\ \hline
\multirow{2}{*}{Chloramphenicol} &  &  0.110 & 0.063 & 0.296 & 0.240 & 0.172
\\
& & (0.072, 0.154) & (-0.097, 0.267) & (0.145, 0.485) & (0.127, 0.374) & (0.119, 0.272)
		\\ \hline
\multirow{2}{*}{Erythromycin}    & 0.011 &  & 0.137 & 0.032 & 0.093 & 0.082 
		\\
		& (0.003, 0.027) && (-0.043, 0.123) & (0.061, 0.238) & (0.026, 0.187) & (0.037, 0.142)
		\\ \hline
\multirow{2}{*}{Penicillin} & 0.002 & 0.007 &  & 0.096 & 0.853 & 0.126 
		\\
		& (-0.023,  0.027) & (-0.006, 0.029) && (-5.122, 4.327) & (-10.192, 9.067) & (-0.356, 0.733)
		\\ \hline
\multirow{2}{*}{Tetracycline}  & 0.003 & 0.002 & 0.001 &  &  0.083 & 0.018
		\\
		& (-0.003, 0.009) & (-0.002, 0.008) & (-0.293, 0.278) && (-0.865, 1.068) & (-1.013, 0.982)
		\\ \hline
\multirow{2}{*}{Co-trimoxazole} & 0.004  & 0.005 & 0.010 & 0.003 &  & 0.177
\\
& (-0.023, 0.028) & (-0.013, 0.020) & (-0.712, 0.714) & (-0.024, 0.030) & & (0.022, 0.453)
		\\ \hline
\multirow{2}{*}{invasiveness} & 0.001 & 0.001 & 0.001 & 0.002 & 0.002 & 
\\
& (-0.027, 0.031) & (-0.016, 0.019) & (-0.023, 0.033) & (-0.029, 0.031) & (-0.002, 0.009) &
\\
\hline\hline
\end{tabular}
}
\label{tb_SPinvasive_appendix}
\end{table}

\begin{table}[!ht]
\footnotesize
\centering
\caption{\textbf{Ecoli data}. genetic covariance in the lower diagonals,  covariance of causal effects in upper diagonals. }
\makebox[\textwidth]{
\begin{tabular}{| l | c| c| c |c| c| c |}
\hline\hline
	& Amoxicillin
	& Cefotaxime
	& Ceftazidime
	& Cefuroxime
	& Ciprofloxacin
	& Gentamicin
\\ \hline
\multirow{2}{*}{Amoxicillin} &  & 0 & 0 & 0 & -0.0002 & 0
\\
& & (0, 0) & (0, 0) & (0, 0) & (-0.007, 0.005) & (0, 0)
		\\ \hline
\multirow{2}{*}{Cefotaxime} & -0.010 &  & 0.046 & 0.032 & -0.013 & 0.007 
		\\
		& (-0.297, 0.246) && (-0.119, 0.244) & (-0.093, 0.156) & (-0.073, 0.040) & (-0.031, 0.038)
		\\ \hline
\multirow{2}{*}{Ceftazidime} & 0.0 & 0.027 & & 0.005 & 0.010 & 0.006
		\\
		& (-0.181, 0.181) & (-2.693, 2.991) & & (-0.046, 0.091) & (-0.037, 0.060) & (-0.016, 0.037)
		\\ \hline
\multirow{2}{*}{Cefuroxime} & -0.009 & 0.041 & 0.022 &  & 0.044 & 0.006
		\\
		& (-0.179, 0.131) & (-1.969, 2.339) & (-1.111, 1.268) & & (-0.059, 0.126) & (-0.015, 0.028)
		\\ \hline
\multirow{2}{*}{Ciprofloxacin} & -0.008 & 0.045 & 0.029 & 0.055 &  & 0.100
\\
& (-0.032, 0.006) & (-0.396, 0.493) & (-0.205, 0.268) & (-0.195, 0.298) &  & (0.046, 0.159) 
		\\ \hline
\multirow{2}{*}{Gentamicin} &  0.0005 &  0.017 & 0.015 &  0.019 & 0.037 & 
\\
& (-0.235, 0.239) & (-3.280, 3.530) & (-1.922, 2.035)  & (-1.997, 2.143)  & (-0.390, 0.474)  & 
\\
\hline\hline
\end{tabular}
}
\label{tb_ecoli_appendix}
\end{table}

\begin{table}
\footnotesize
\centering
\caption{ \textbf{Neisseria gonorrhoeae data}. genetic covariance in the lower diagonals,  covariance of causal effects in upper diagonals.}
\begin{tabular}{| l | c |c |c |c |c | }
\hline\hline
	& Azithromycin
	& Cefixime
	& Ciprofloxacin
	& Penicillin
	& Tetracycline
\\ \hline
\multirow{2}{*}{Azithromycin} &  & 0 & 0 & -0.040 & 0
\\
& & (0, 0) & (0, 0) & (-1.834, 1.641)& (0, 0)
		\\ \hline
\multirow{2}{*}{Cefixime}  & -0.019 &  & -0.507 & 0.106 & 0
		\\
		& (-0.175, 0.184) &  & (-2.413, 1.125) & (-0.059, 0.487)  & (0, 0)
		\\ \hline
\multirow{2}{*}{Ciprofloxacin} &  -0.865 & -0.173 &  & 0.047 & 0.301
		\\
		& (-5.776, 2.457) & (-0.904, 0.341)  &  & (-0.194, 0.308) & (0.162, 0.480)
		\\ \hline
\multirow{2}{*}{Penicillin}  & -0.009 & 0.028 & 2.259 &  & -0.037
		\\
		& (-6.299, 5.817) & (-0.408, 0.351) & (-2.824, 10.803) &  & (-0.466, 0.402)
		\\ \hline
\multirow{2}{*}{Tetracycline} & -0.153 & 0.004 & 1.469 &  0.482 & 
\\
& (-0.725, 0.146) & (-0.187, 0.243) & (0.163, 2.770) & (0.205, 0.832) &
		\\ \hline\hline
\end{tabular}
\label{tb_NG_appendx}
\end{table}

\end{document}